# The Role of Instructional Guidance in Generative AI-Assisted Learning: Empirical Evidence from Construction Engineering Education

Xiaoyu Hou[1], Bo Xiao[2], Hexu Liu[3] and Shane Mueller[4]

[1]PhD Student, Dept. of Civil, Environmental, and Geospatial Engineering, Michigan Technological Univ., Houghton, MI. 49931, U.S.A. Email: xiaoyuho@mtu.edu

[2]Assistant Professor, Dept. of Civil, Environmental, and Geospatial Engineering, Michigan Technological Univ., Houghton, MI. 49931 Email: boxiao@mtu.edu (Corresponding Author)

[3]Associate Professor, Dept. of Civil and Construction Engineering, Western Michigan Univ., Kalamazoo, MI. 49008 Email: hexu.liu@wmich.edu

[4]Professor, Dept. of Psychology and Human Factors, Michigan Technological Univ., Houghton, MI. 49931 Email: shanem@mtu.edu

**ABSTRACT:**

Generative artificial intelligence (AI) is increasingly used to support self-directed learning, yet student interaction with such systems often remains unstructured, limiting engagement in deeper cognitive processes. This study examines how instructional guidance shapes student–AI interaction in construction education. A five-step prompting framework grounded in Generative Learning Theory (GLT) is introduced to guide learner interaction during review activities. A controlled experiment compares three learning conditions: slide-based learning, unprompted AI-supported learning, and prompted AI-supported learning. Learning performance is assessed using multiple-choice and open-ended tasks, and user experience is measured using the User Experience Questionnaire (UEQ). Performance differences are concentrated on tasks requiring explanation and reasoning. The prompted condition achieves higher open-ended scores, with an improvement of approximately 2–3 points on a scale of 18 ($p < 0.01$), while no significant differences are observed in multiple-choice performance. The unprompted condition remains comparable to slide-based learning. These findings indicate that the effectiveness of AI-supported learning depends on how interaction is structured. The proposed framework provides a basis for integrating learning science principles into generative AI systems for construction education.

**Keywords:** Construction Education, Generative artificial intelligence, Empirical study, AI-supported learning, Self-directed learning

## INTRODUCTION

Generative artificial intelligence (AI) is increasingly being incorporated into educational settings as a tool for learning support and instructional assistance (AlAli and Wardat 2024). A substantial body of empirical research has examined generative AI–based systems as tools that extend learning beyond traditional classroom instruction, particularly during independent and self-directed study. Prior studies have explored applications such as on-demand explanation generation, interactive question answering, and adaptive support that respond to learners' evolving needs during review activities (Bernasconi, Redavid and Ferilli 2025; Leon 2024; Yogi, Chowdary and Santhoshi 2024). Experimental research on intelligent tutoring systems and conversational agents has shown that AI-generated explanations can assist learners in articulating conceptual understanding, revisiting previously learned material, and reflecting on problem-solving strategies during independent study (Aluko et al. 2025; El Fathi et al. 2025; Mueller et al. 2025). Other controlled studies have further reported that generative AI–supported learning environments can facilitate formative feedback and sustain learner engagement when students interact with conceptually complex or information-dense instructional content outside scheduled class time (Wu and Chiu 2025; Zhan et al. 2025). Collectively, this body of work positions generative AI as a learning support that shapes how learners engage with explanation, feedback, and self-directed study activities.

Within construction education, generative artificial intelligence has increasingly been explored as a supplementary learning resource. Existing studies have examined the use of conversational generative AI tools to support students' engagement with course content, generate explanations, and provide feedback during independent or course-related learning activities (Oblitas, Adeeb and Cruz-Noguez 2025; Uddin et al. 2024). Additional studies have shown that large language model–based systems can assist students in clarifying conceptual understanding and organizing solution approaches in engineering contexts, particularly when tasks involve interpretation rather than advanced analytical computation (Matzakos and Moundridou 2025). At the instructional system level, research has primarily emphasized the design and implementation of AI-enabled platforms, including approaches that align AI-generated responses with construction-specific knowledge bases and course materials. Prior work by the authors followed this trajectory by developing a retrieval-augmented generative AI platform for construction education and demonstrating its technical feasibility and instructional alignment through expert-based and formative evaluation (Hou et al. 2026). These studies establish that generative AI can be integrated into construction education settings and can function as a viable learning support at the application and system levels.

Despite these developments, student interaction with generative AI in construction education is often implemented as an open-ended activity, in which learners independently formulate questions without structured guidance on how to engage in cognitively effective learning processes. From a learning science perspective, meaningful learning depends not only on access to explanations but also on whether learners are prompted to actively construct understanding through processes such as explanation, organization, and integration. When AI-supported learning remains unstructured, students may rely on AI-generated responses for surface-level clarification without engaging in deeper generative processing. Although generative AI tools are increasingly available in construction education, empirical evidence remains limited regarding how instructional guidance shapes students' interaction with generative AI and influences learning processes. Controlled comparisons between guided AI interaction, unprompted AI interaction, and non-AI learning conditions remain underexplored.

Building on prior work in which a course-aligned generative AI platform was developed for construction education contexts (Hou et al. 2026), the present study shifts the focus from system feasibility to the pedagogical design of student–AI interaction during self-directed learning. Specifically, this study introduces a structured interaction approach, operationalized through a five-step prompting sequence grounded in Generative Learning Theory (GLT), to guide how students engage with AI while learning course content (Mayer 2009). A controlled experiment was conducted under a shared instructional input with three learning conditions: conventional slide-based learning, unprompted AI-supported learning, and prompted AI-supported learning guided by the proposed five-step framework. By comparing learning outcomes and user experience across the three conditions, the study examines whether theory-guided prompt produces measurable instructional benefits and provides evidence to support more principled integration of generative AI in construction education. To guide the investigation, the study addresses the following research questions:

RQ1: Does instructional guidance improve higher-order learning outcomes compared with unprompted AI-supported and slide-based learning?

RQ2: Does instructional guidance influence user experience in AI-supported learning environments?

## LITERATURE REVIEW

This chapter reviews prior work from three complementary perspectives: generative artificial intelligence applications in construction education, structured cognitive support in digital learning environments, and the theoretical foundations provided by GLT.

### Generative AI–Supported Learning in Construction Education

Generative AI, particularly large language models (LLMs), is increasingly used in construction and civil engineering education to provide conversational explanation, responsive feedback, and adaptive learning support. Current literature consistently characterizes generative AI as a supplementary instructional resource whose educational value depends on alignment with course objectives and instructional design, rather than as an autonomous teaching agent (Jelodar 2025). Empirical studies indicate that LLMs are most effective in facilitating conceptual understanding and interpretive reasoning, while their performance remains limited in advanced analytical and calculation-intensive problem-solving tasks (Gesnot 2025). Exam-based evaluations show that conversational AI can assist learners in organizing solution procedures and clarifying underlying engineering concepts, but accuracy declines in tasks requiring domain-specific computation or highly specialized expertise (Oblitas, Adeeb and Cruz-Noguez 2025). Comparable findings are reported in laboratory-based instruction, where LLMs integrated alongside formal mathematical tools improve methodological transparency and conceptual comprehension without eliminating the need for disciplinary verification and instructor oversight (Matzakos and Moundridou 2025).

Learning outcomes associated with generative AI are strongly influenced by how students are guided to interact with these systems. Research on interaction design demonstrates that structured prompting strategies can significantly improve LLM performance on previously incorrect engineering exam questions, underscoring the role of students’ inquiry practices in mediating educational benefit (Genç 2025). At the same time, classroom evidence shows that while AI-assisted explanations can enhance the clarity and depth of student responses, unstructured use often results in superficial engagement, reinforcing the need for instructional scaffolding that emphasizes critical evaluation over answer acceptance (Uddin et al. 2024).

At the course and curriculum levels, generative AI has been incorporated into task-oriented and project-based learning environments, where LLMs support open-ended activities such as programming for construction workflows and reasoning in project management contexts when embedded within problem-based instructional designs (Maalek 2024). In contrast, assessment-oriented applications reveal substantive limitations. Studies evaluating LLM use in grading complex student work, including construction management capstone projects, report limited alignment between AI-generated assessments and expert judgment, particularly for subjective and interpretive criteria (Castelblanco, Cruz-Castro and Yang 2024). More recent research extends beyond activity-level support toward adaptive and personalized learning frameworks, combining LLM-based conversational agents with learner modeling techniques to support progressive mastery of procedural skills across academic and professional training settings (Mueller et al. 2025). In construction education, generative AI demonstrates educational value primarily when embedded within pedagogically structured learning environments, while uncritical or assessment-substitutive use remains constrained by reliability and validity concerns (Jelodar 2025).

Research on generative AI in construction and civil engineering education has largely focused on LLM-supported learning for conceptual clarification, guided reasoning, and project-based instructional activities. These studies report pedagogical benefits under structured prompting and instructional scaffolding, while also noting persistent limitations in analytical reliability and assessment validity. However, empirical evidence remains limited regarding how generative AI-supported learning influences student engagement, learning processes, and instructional effectiveness within authentic construction education contexts, leaving a gap between exploratory instructional applications and systematically evaluated learning outcomes.

**Structured Cognitive Support in Digital Learning**

Research in technology-enhanced learning has extensively examined how learners regulate cognitive and metacognitive processes during self-directed study in computer-based environments (Persico and Steffens 2017; Sui, Yen and Chang 2024). Studies conducted in hypermedia and related digital settings report substantial variability in learners' planning, monitoring, and evaluative behaviors, and many students demonstrate limited spontaneous use of self-regulatory strategies when instructional support is minimal (Azevedo and Cromley 2004; Bannert 2006). In response, structured cognitive and metacognitive supports, commonly implemented through prompts that activate reflection and regulation, have been systematically investigated as instructional design mechanisms to promote deeper processing and improve learning outcomes (Hsu, Wang and Zhang 2017; Li et al. 2025).

A substantial empirical literature evaluates the effects of such prompts in digital learning contexts. A meta-analysis of experimental studies in computer-based environments found that metacognitive prompts significantly improved both self-regulated learning activities ($g = 0.50$) and learning outcomes ($g = 0.40$), providing quantitative support for structured prompting as an effective instructional intervention (Guo 2022). Experimental and quasi-experimental research further indicates that prompts encouraging planning, monitoring, and evaluation enhance strategic engagement and, under specific instructional conditions, improve learning performance and self-efficacy (Gentner 2022). Sequential analyses of inquiry-based learning tasks show that structured prompts influence learners' regulatory patterns, with more successful learners demonstrating systematic monitoring and evaluation during prompted activities (Hsu, Wang and Zhang 2017). Additional evidence suggests that prompt effectiveness depends on design features such as specificity, feedback integration, and adaptability (Engelmann, Bannert and Melzner 2021). These findings establish structured prompting as a validated mechanism for supporting generative and self-regulated learning in digital environments.

Most investigations of structured cognitive support, however, have been conducted in hypermedia systems, inquiry-based platforms, or intelligent tutoring environments. Generative AI introduces a different interaction format based on conversational exchange, where learners often initiate open-ended questions without structured guidance. Whether the documented benefits of prompting in prior digital settings extend to generative AI–supported learning remains unclear. Empirical examination is therefore needed to determine how structured cognitive support functions within conversational AI environments, particularly in domain-specific contexts such as construction education.

**Generative Learning Theory**

Generative Learning Theory conceptualizes learning as a constructive process in which learners organize information and relate it to prior knowledge to build coherent mental representations (Wittrock 1989). Within this perspective, understanding emerges through cognitive activity that transforms content during learning. Core generative processes include selecting relevant information, organizing relationships among ideas, and integrating new material with existing schemas (Fiorella and Mayer 2015) (Lee, Lim and Grabowski 2008). These processes collectively support the formation of structured knowledge representations that enable transfer and flexible application. Conceptually, these generative transformations can be understood as progressing from initial clarification of ideas to structural organization, and ultimately to integrative refinement of understanding.

Empirical research has examined how generative activities function in digital and multimedia environments. A meta-analysis of self-explanation studies reported positive effects on conceptual understanding and transfer, with outcomes influenced by the structure and quality of the prompts provided (Bisra et al. 2018). Reviews of generative strategies in higher education indicate that summarizing, drawing, mapping, and self-testing are effective when tasks require learners to reorganize and integrate ideas (Fiorella and Mayer 2016). Experimental studies in multimedia contexts report comparable findings. Interactive graphic organizers are associated with deeper integrative processing and improved comprehension compared with static text or completed representations (Wang et al. 2021). Research applying segmenting and signaling principles to organizer design also reports gains in retention, transfer, and cognitive efficiency (Qin and Wu 2025). These studies consistently emphasize the role of learner-generated structuring and integration in supporting durable understanding.

Generative processing is unevenly enacted in open digital learning environments. In self-directed settings, learners frequently focus on information access or solution completion, while organization and conceptual integration remain implicit rather than deliberate. Although generative learning principles have informed multimedia and inquiry-based instructional design, their integration into conversational generative AI contexts has received limited systematic attention. Interaction with generative AI systems typically centers on explanation delivery. Whether such exchanges consistently elicit the organizing and integrative processes described in generative learning theory remains empirically unresolved.

Existing work suggests that generative AI can support learners and that structured prompts can improve digital learning, but there is limited experimental evidence in construction education comparing guided AI interaction, unguided AI interaction, and conventional review within a common instructional setting.

## METHODOLOGY

### Overall Research Design and Framework

The study adopted a randomized controlled design to examine how different forms of student–AI interaction influence learning during self-directed review. To ensure comparability across groups, all participants first completed a common instructional phase using identical PPT-based materials that introduced foundational construction concepts. This shared baseline established equivalent prior exposure before the intervention stage.

Following the baseline phase, participants were randomly assigned to one of three learning conditions that differed only in the structure of post-baseline interaction. In the slide-based continuation condition, students reviewed static instructional materials without adaptive feedback. In the unprompted AI condition, students engaged freely with a retrieval-augmented generation (RAG)-based generative AI system. In the prompted AI condition, students interacted with the same AI system but followed a predefined five-step sequence derived from generative learning principles to guide clarification, organization, and integration processes during review.

By holding instructional input constant and varying only the interaction format, the design isolates the role of structured prompting in shaping learning outcomes and user experience. Figure 1 presents the overall experimental framework.

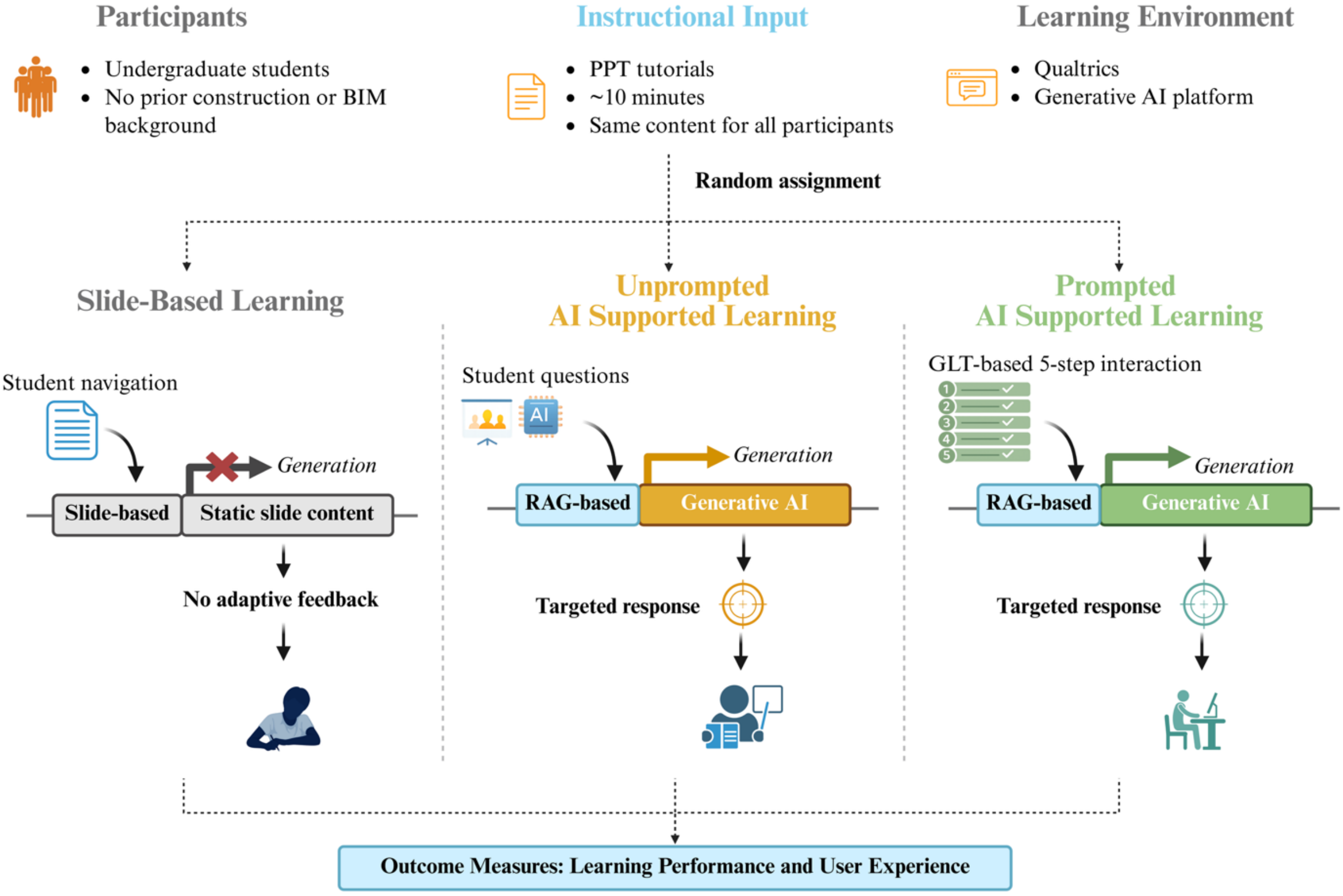

**Fig. 1.** Overall research framework and two-study design

## AI Platform

The study was conducted using a web-based generative AI–supported learning platform developed for construction education (Figure 2). The platform enables document-grounded interaction with course materials and generates source-linked responses aligned with curated instructional resources. It adopts a retrieval-augmented generation (RAG) architecture that integrates semantic retrieval with large language model generation, thereby constraining responses to course-specific content.

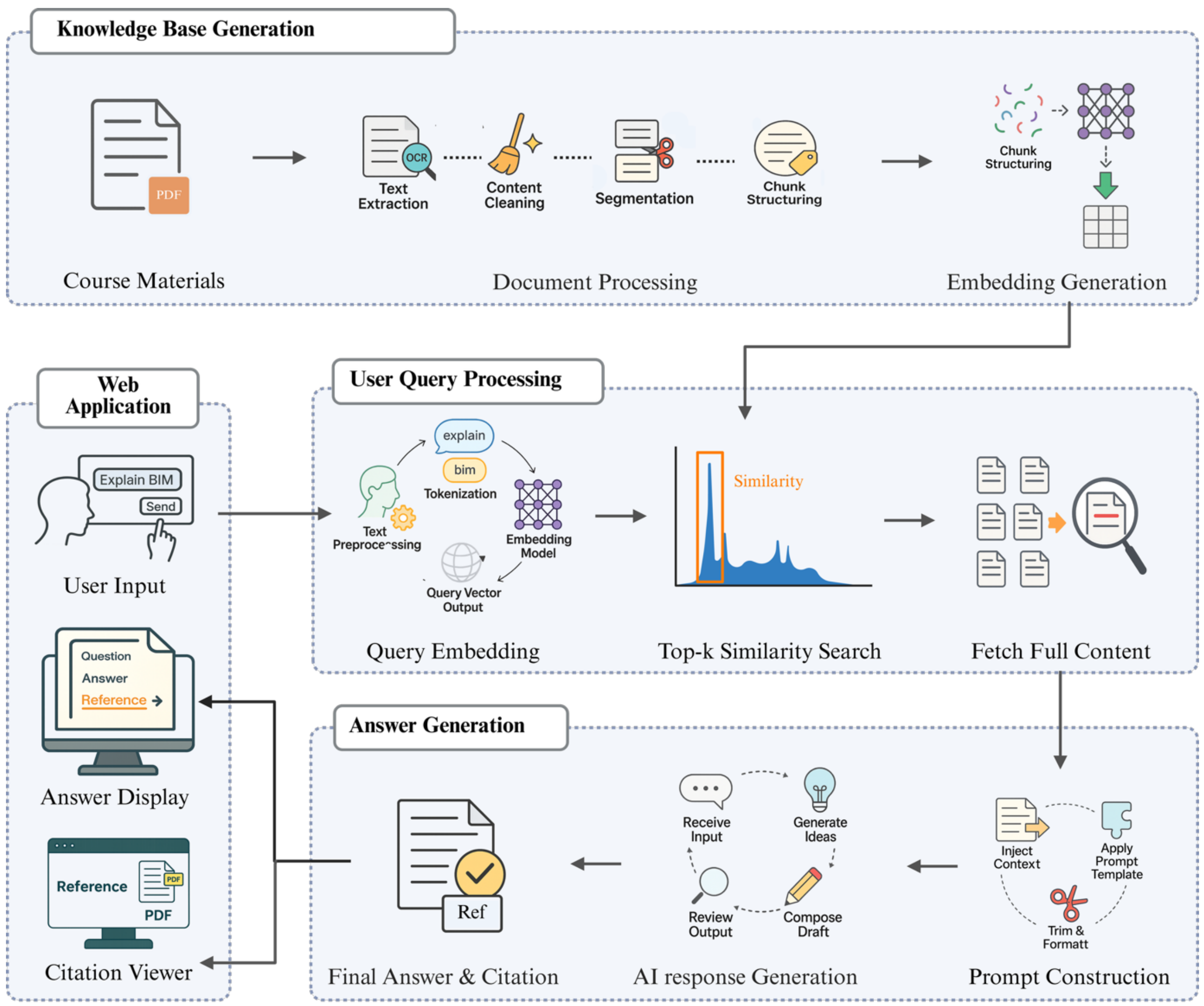

**Fig. 2.** Overview of the generative AI–based learning platform used in this study (Hou et al. 2026).png

Prior evaluations established the platform's reliability and instructional alignment. These included expert review of response accuracy and relevance, automated assessment of grounding and completeness, and a formative classroom evaluation of usability and interaction patterns. Together, these stages provided a stable baseline suitable for experimental use. In the present experiment, the same platform configuration and course-aligned knowledge base were used across both AI-supported conditions. The unprompted and prompted groups interacted with an identical system and accessed the same instructional corpus.

**Prompting framework and theoretical grounding**

The prompted AI condition was structured using a five-step interaction framework grounded in GLT. GLT conceptualizes learning as an active process in which learners construct meaning by selecting relevant information, organizing key elements, and integrating new material with prior knowledge. The five-step prompting framework was designed to operationalize these generative processes within AI-supported review. The framework guides students through a structured sequence consisting of Clarify, Organize, Integrate, Differentiate, and Correct. Each step corresponds to a distinct cognitive function aligned with the selecting, organizing, and integrating mechanisms described in GLT.

The five-step prompting framework was derived by translating the core generative processes described in GLT into observable interaction behaviors within an AI-supported environment. As illustrated in Figure 3, the resulting sequence structures student–AI interaction as a progressive cognitive process unfolding through five interconnected stages. The operationalization of the five-step prompting framework, including the specific prompts used in the experiment and their alignment with GLT, is summarized in Table 1.

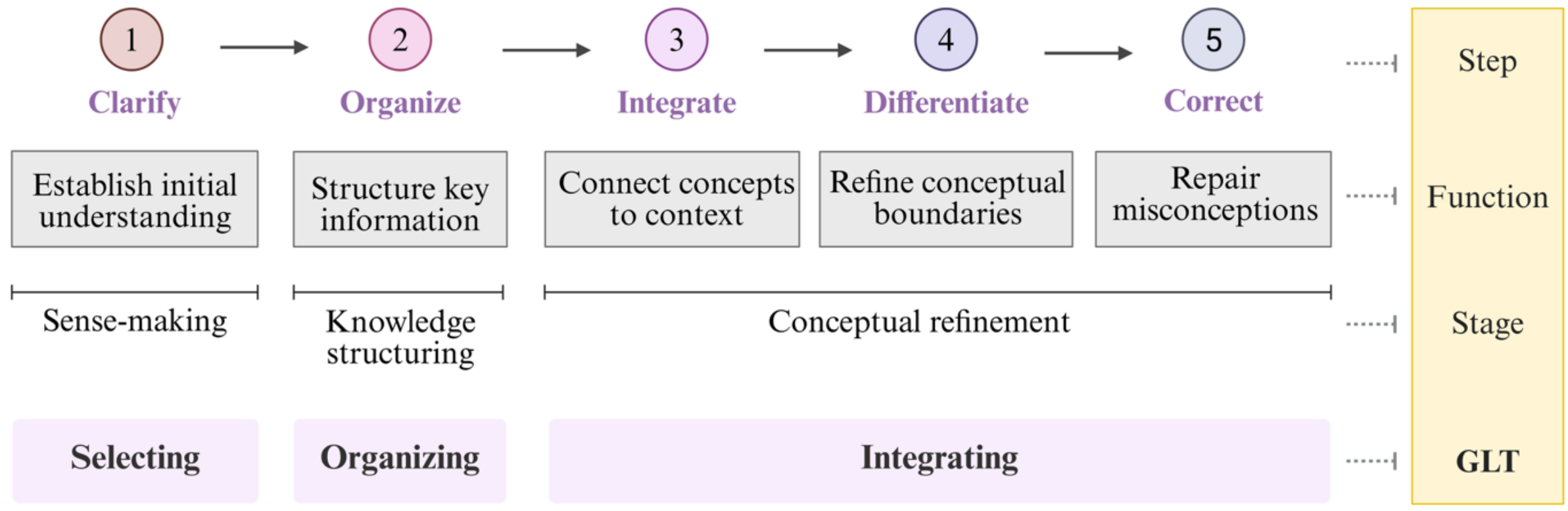

**Fig. 3.** Five-step prompting framework aligned with Generative Learning Theory

The sequence begins with (1) Clarify, which directs attention to core concepts and supports the establishment of an initial understanding. Students request focused explanations or articulate preliminary interpretations, thereby activating key conceptual elements relevant to the topic. Attention then shifts in (2) Organize toward arranging selected information. Learners identify relationships among components, outline central ideas, and construct an internal structure that transforms dispersed details into coherent representations.

**Table 1.** Operationalization of the five-step prompting framework aligned with GLT

| Step | GLT function | Instructional purpose | Example student prompt | Intended cognitive action |
|---|---|---|---|---|
| Clarify | Selecting | Direct attention to the core meaning of a concept | "Explain [X] in 2–3 simple sentences, focusing only on the core meaning." | Identify key ideas and establish initial understanding |
| Organize | Organizing | Highlight key points within the concept | "Give me the 2–3 most important key points of [X], listed in order of importance." | Prioritize important information and structure understanding |
| Integrate | Integrating | Connect the concept to practical contexts | "Give me a real-world example of [X] that clearly includes its key features: [A], [B], and [C]." | Link conceptual knowledge with real-world application |
| Differentiate | Integrating (refinement) | Distinguish related or easily confused concepts | "Identify a concept that students often confuse with [X] and explain the key differences between them." | Compare concepts and refine conceptual boundaries |
| Correct | Integrating (evaluation) | Address misconceptions and reinforce accurate understanding | "What is the most important misconception about [X], and why is it incorrect?" | Evaluate understanding and correct misunderstandings |

This structured understanding is extended in (3) Integrate, where students connect concepts across contexts, relate definitions to practical scenarios, and link new material with prior knowledge, allowing meaning to develop beyond isolated recall. The subsequent stage, (4) Differentiate, refines conceptual boundaries by encouraging comparison among closely related terms, examination of procedural distinctions, and consideration of limiting conditions, reducing ambiguity and increasing precision. The process concludes with (5) Correct, during which students re-examine interpretations against cited sources, identify inconsistencies, and revise misunderstandings, consolidating conceptual accuracy and stability.

Across the five steps, the interaction sequence structures AI use as a guided cognitive activity rather than an open-ended query process. The progression from clarification to correction mirrors the generative learning cycle, moving from initial sense-making through organization and integration toward conceptual refinement. Figure 3 illustrates the alignment between the five-step interaction framework and the cognitive stages described in GLT.

**Experimental Procedure**

The experiment is implemented as an online controlled study using the Qualtrics platform. Participants are undergraduate students enrolled at Michigan Technological University. Recruitment is restricted to students without prior coursework or practical experience in construction management, BIM, or closely related domains to minimize variation in baseline familiarity with the instructional content. The study protocol is reviewed and approved by the Institutional Review Board, and all participants provide electronic informed consent prior to participation. A total of 119 participants were initially recruited and assigned to three learning conditions, including 45 in the slide-based condition (Control), 37 in the prompted AI-supported condition, and 37 in the unprompted AI-supported condition. Prior to analysis, responses were screened for completeness. Cases with missing data in either the UEQ or learning performance measures were removed, as incomplete records would affect the calculation of composite scores. Following this screening process, 24 responses were excluded, resulting in a final sample of 95 participants. The retained sample included 33 participants in the slide-based condition, 29 in the prompted condition, and 33 in the unprompted condition. Exclusion counts were 12, 8, and 4 for the three conditions, respectively.

All participants complete the study individually and follow an identical sequence of activities during the initial phase. The session begins with an introduction and consent procedure, followed by a standardized baseline learning phase. During this phase, all participants review the same PPT-based instructional materials covering introductory construction-related concepts. The materials are presented in a fixed format to ensure equivalent exposure across groups and to establish a common knowledge baseline prior to experimental manipulation. Upon completion of the baseline phase, participants are randomly assigned to one of three post-baseline learning conditions: (1) continued slide-based review (Control condition), (2) unprompted AI-supported review, or (3) structured prompted AI-supported review.

In the control group, participants continue reviewing the same PPT materials independently without access to AI support. In the unprompted AI condition, participants engage with the course-aligned retrieval-augmented generative AI platform described in Section 3.2. They are free to pose questions in natural language and interact with the system without guidance regarding question formulation or sequencing. In the prompted AI condition, participants interact with the same AI platform but follow the predefined five-step interaction framework grounded in GLT. They apply the full sequence of prompts for each selected topic before proceeding to another topic. The instructional corpus, system configuration, and review duration remain constant across conditions to isolate the effect of interaction structure.

Following the review phase, all participants complete the same outcome measures. Learning performance is assessed through a post-intervention test consisting of multiple-choice items and open-ended explanation questions aligned with the instructional materials. Open-ended responses are evaluated using analytic scoring rubric. Two trained raters independently score the responses, and discrepancies are resolved through discussion. User experience is measured using the short version of the User Experience Questionnaire (UEQ) (Schrepp et al. 2017). Figure 4 illustrates the implementation of the experimental procedure in Qualtrics, including the key survey blocks and the condition-specific branches.

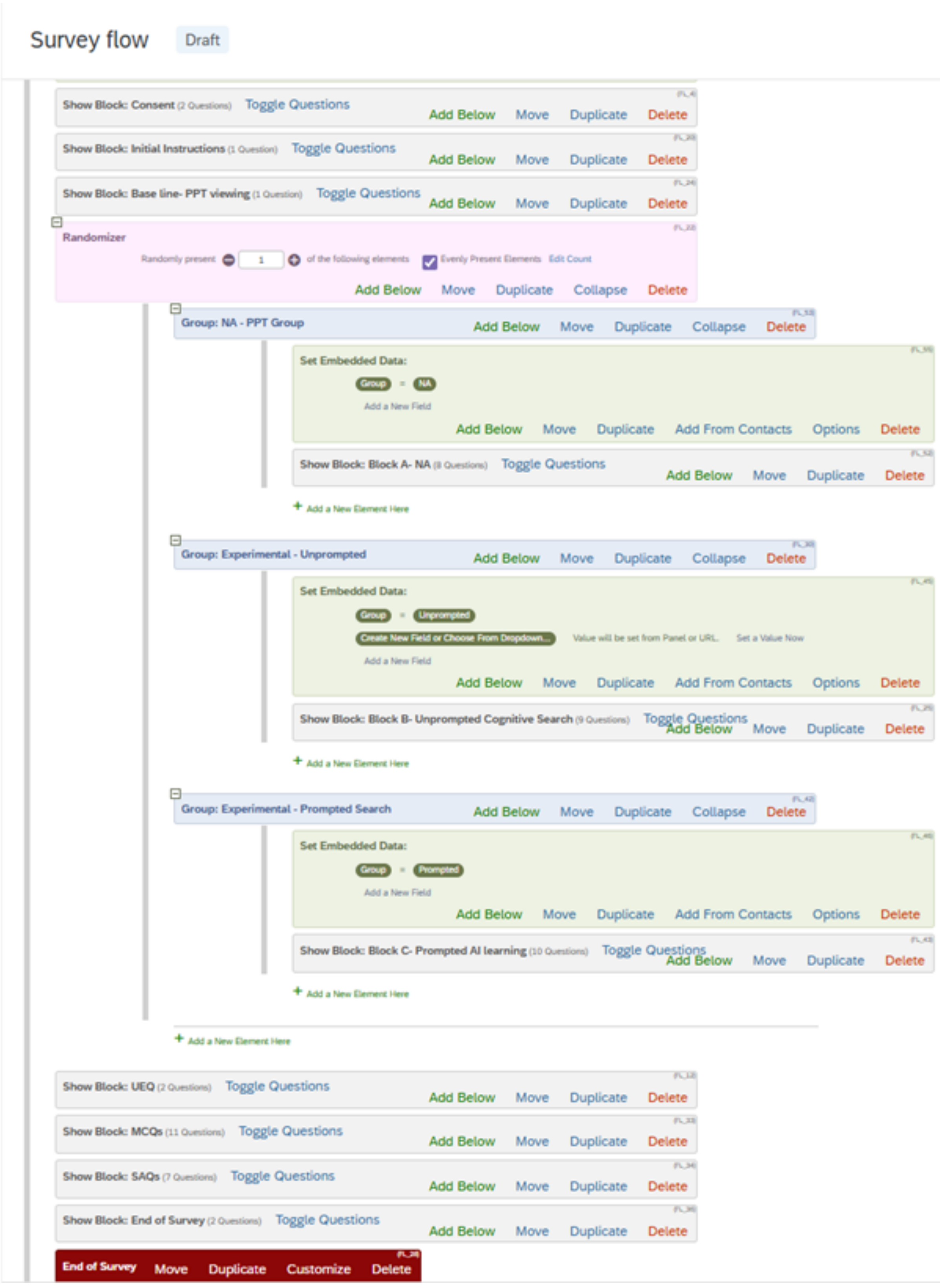

**Fig. 4.** Implementation of the experimental procedure in Qualtrics

### Data Collection and analysis

Data collection focuses on two dimensions of instructional effectiveness: learning performance and learning experience. A common set of outcome measures is applied across all conditions to ensure comparability.

Learning performance is assessed through a post-intervention test aligned with the instructional materials. The assessment includes both multiple-choice and open-ended items, capturing different levels of cognitive processing. Multiple-choice questions are developed directly from the course materials to ensure content validity. Open-ended responses are evaluated using an analytic scoring rubric emphasizing conceptual accuracy, completeness, and clarity. All responses are independently scored by two raters who are blind to experimental condition. Inter-rater reliability is assessed using the intraclass correlation coefficient (ICC) based on a two-way random-effects model for absolute agreement (Shrout and Fleiss 1979), indicating excellent agreement between raters (ICC = 0.91). Final scores are determined by averaging the two ratings.

Learning experience is measured using the User Experience Questionnaire (UEQ), which consists of 26 items grouped into six dimensions: Attractiveness, Perspicuity, Efficiency, Dependability, Stimulation, and Novelty (Schrepp et al. 2017). Responses are recorded on a seven-point bipolar scale and transformed to a standardized range from −3 to +3 following UEQ guidelines. Item scores are polarity-adjusted and aggregated to compute mean values for the predefined pragmatic and hedonic dimensions.

Group differences in learning performance and user experience are examined using one-way analysis of variance (ANOVA). Assumptions of normality and homogeneity of variance are verified prior to analysis. Statistical significance is evaluated at the conventional alpha level of 0.05.

## EXPERIMENT RESULTS

### User experience results

User experience was assessed using the short version of the UEQ, which consists of 26 semantic differential items organized into six predefined dimensions: Attractiveness, Perspicuity, Efficiency, Dependability, Stimulation, and Novelty. These dimensions capture complementary aspects of interaction quality, with Perspicuity, Efficiency, and Dependability reflecting pragmatic characteristics related to task-oriented use, and Stimulation and Novelty reflecting hedonic characteristics associated with engagement and perceived innovativeness. Attractiveness functions as an overall evaluative dimension integrating both pragmatic and hedonic impressions. Item responses were polarity-corrected and centered to yield scores on a standardized scale ranging from −3 to +3.

Following preprocessing, the UEQ results are summarized in Table 2, and the dimension-level comparisons across the three learning conditions are presented in Figure 5. The findings reveal a differentiated pattern between pragmatic and hedonic dimensions.

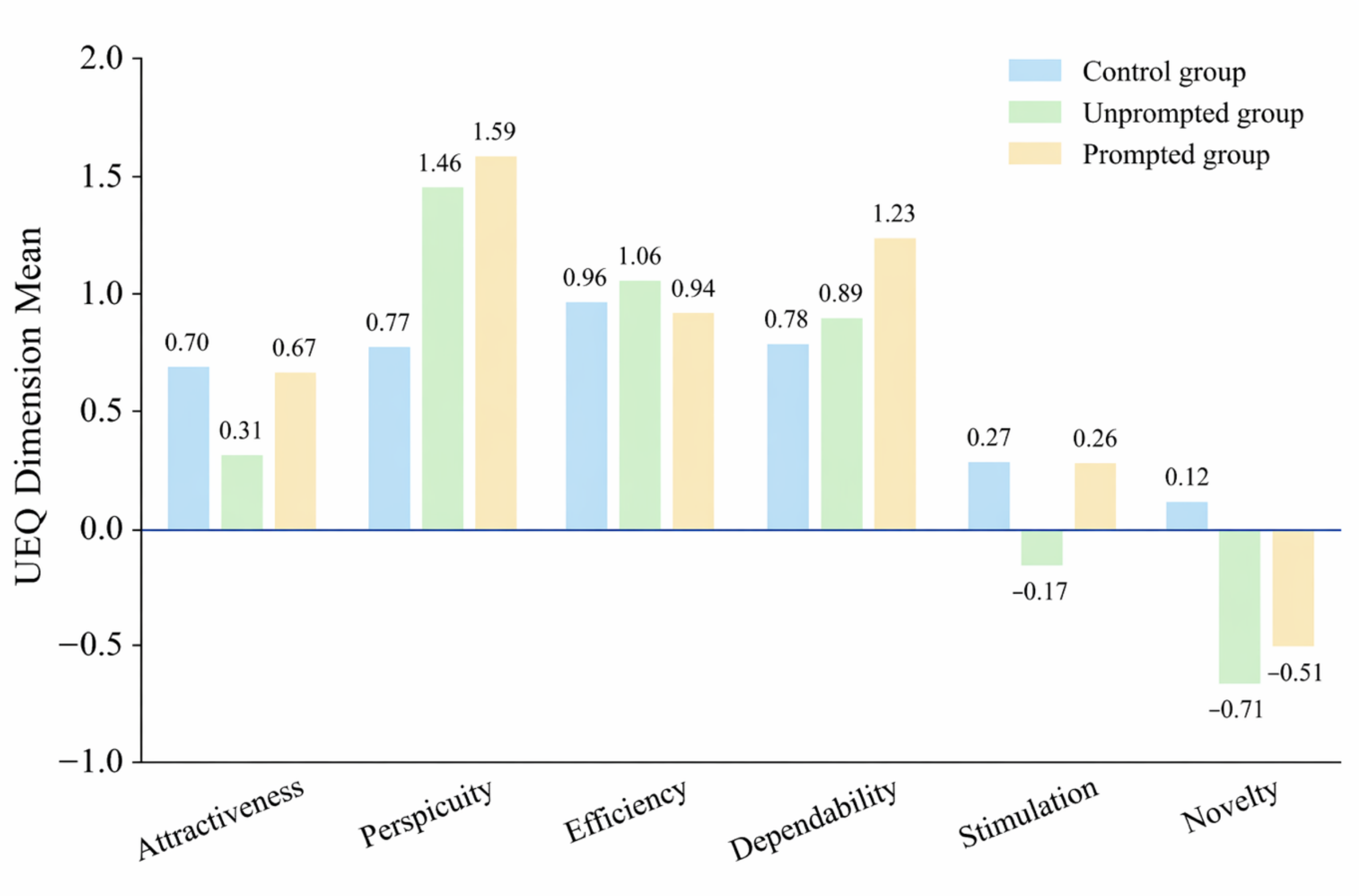

**Fig. 5.** UEQ dimension scores across three learning conditions

**Table 2.** UEQ dimension scores across three learning conditions

| Item | Control group | Unprompted group | Prompted group |
|---|---|---|---|
| **Attractiveness** | | | |
| Q1 | 0.30 | 0.36 | 0.69 |
| Q12 | 1.09 | 0.58 | 1.10 |
| Q14 | 0.52 | 0.24 | 0.59 |
| Q16 | 0.58 | 0.09 | 0.90 |
| Q24 | 0.67 | -0.30 | -0.28 |
| Q25 | 1.03 | 0.88 | 1.03 |
| Dimension Mean | **0.70** | **0.31** | **0.67** |
| **Perspicuity** | | | |
| Q2 | 0.97 | 1.45 | 1.59 |
| Q4 | 0.45 | 1.39 | 1.14 |
| Q13 | 0.52 | 1.42 | 1.93 |
| Q21 | 1.12 | 1.58 | 1.69 |
| Dimension Mean | **0.77** | **1.46** | **1.59** |
| **Efficiency** | | | |
| Q9 | 0.67 | 0.33 | 0.17 |

| Item | Control group | Unprompted group | Prompted group |
|---|---|---|---|
| Q20 | 0.88 | 1.03 | 0.97 |
| Q22 | 1.00 | 1.12 | 0.97 |
| Q23 | 1.30 | 1.76 | 1.66 |
| Dimension Mean | **0.96** | **1.06** | **0.94** |
| **Dependability** | | | |
| Q8 | 0.70 | 1.06 | 1.41 |
| Q11 | 0.79 | 0.64 | 1.34 |
| Q17 | 0.61 | 0.48 | 0.86 |
| Q19 | 1.00 | 1.36 | 1.31 |
| Dimension Mean | **0.78** | **0.89** | **1.23** |
| **Stimulation** | | | |
| Q5 | 0.97 | 0.39 | 0.66 |
| Q6 | -0.45 | -0.79 | -0.38 |
| Q7 | 0.00 | -0.42 | 0.24 |
| Q18 | 0.55 | 0.15 | 0.52 |
| Dimension Mean | **0.27** | **-0.17** | **0.26** |
| **Novelty** | | | |
| Q3 | 0.12 | -0.91 | -0.41 |
| Q10 | 0.12 | -0.73 | -0.66 |
| Q15 | -0.12 | -0.97 | -0.76 |
| Q26 | 0.39 | -0.24 | -0.24 |
| Dimension Mean | **0.12** | **-0.71** | **-0.51** |

Note: All scores were transformed by centering the original 1–7 scale to a −3 to +3 range, following the standard UEQ methodology.

Overall, the prompted AI-supported learning condition achieves the highest scores in most pragmatic dimensions, particularly in Perspicuity and Dependability. As shown in Table 2, Perspicuity increases substantially in the Prompted group (1.59) compared to the Control group (0.77), indicating that the structured prompting strategy significantly enhances interaction clarity and reduces cognitive ambiguity. This pattern is also reflected in Figure 5, where the Prompted condition consistently outperforms the other two groups on task-oriented usability dimensions. Similarly, the higher Dependability score observed in the Prompted condition (1.23) suggests that students perceive AI interactions as more predictable and reliable when guided by the five-step prompting framework.

In contrast, Efficiency remains relatively stable across all three conditions, with only marginal variation (0.96, 0.94, and 1.06 for Control, Prompted, and Unprompted, respectively, as shown in Table 2). Although prompting improves interaction clarity, it does not reduce perceived effort or time expenditure. Figure 5 further confirms the absence of a clear advantage for the Prompted group in this dimension, implying that the additional cognitive steps introduced by structured prompting may offset potential efficiency gains.

A different pattern emerges in the hedonic dimensions. Both Stimulation and Novelty exhibit relatively low or negative scores, particularly in AI-supported conditions. As reported in Table 2, Novelty declines from 0.12 in the Control condition to −0.51 and −0.71 in the Prompted and Unprompted conditions, respectively. These results suggest that while AI introduces new interaction mechanisms, structured prompting may reduce perceived novelty and constrain exploratory engagement. Similarly, Stimulation scores remain low across all conditions, indicating that the learning activity is perceived as primarily task-oriented rather than intrinsically engaging.

The Attractiveness dimension, which integrates both pragmatic and hedonic qualities, does not demonstrate a clear advantage for the Prompted condition. Although the Prompted group performs comparably to the Control group, it does not significantly exceed it. This suggests that improvements in specific usability-related dimensions do not necessarily translate into a stronger overall affective evaluation.

From an instructional perspective, these findings indicate that theory-guided prompting enhances the functional quality of student–AI interaction, even if it does not improve perceived enjoyment or novelty. This distinction is particularly relevant in construction education contexts, where effective learning may depend more on cognitive support and clarity than on experiential engagement. This distinction provides a basis for interpreting the learning performance results, particularly in understanding how improvements in clarity and reliability may translate into measurable gains in knowledge acquisition.

**Learning performance results**

Learning performance outcomes across the three conditions are summarized in Table 3 and further visualized in Figure 6, including multiple-choice scores, open-ended responses, and overall performance. The results indicate a consistent advantage for the prompted AI-supported learning condition, particularly in higher-order assessment tasks.

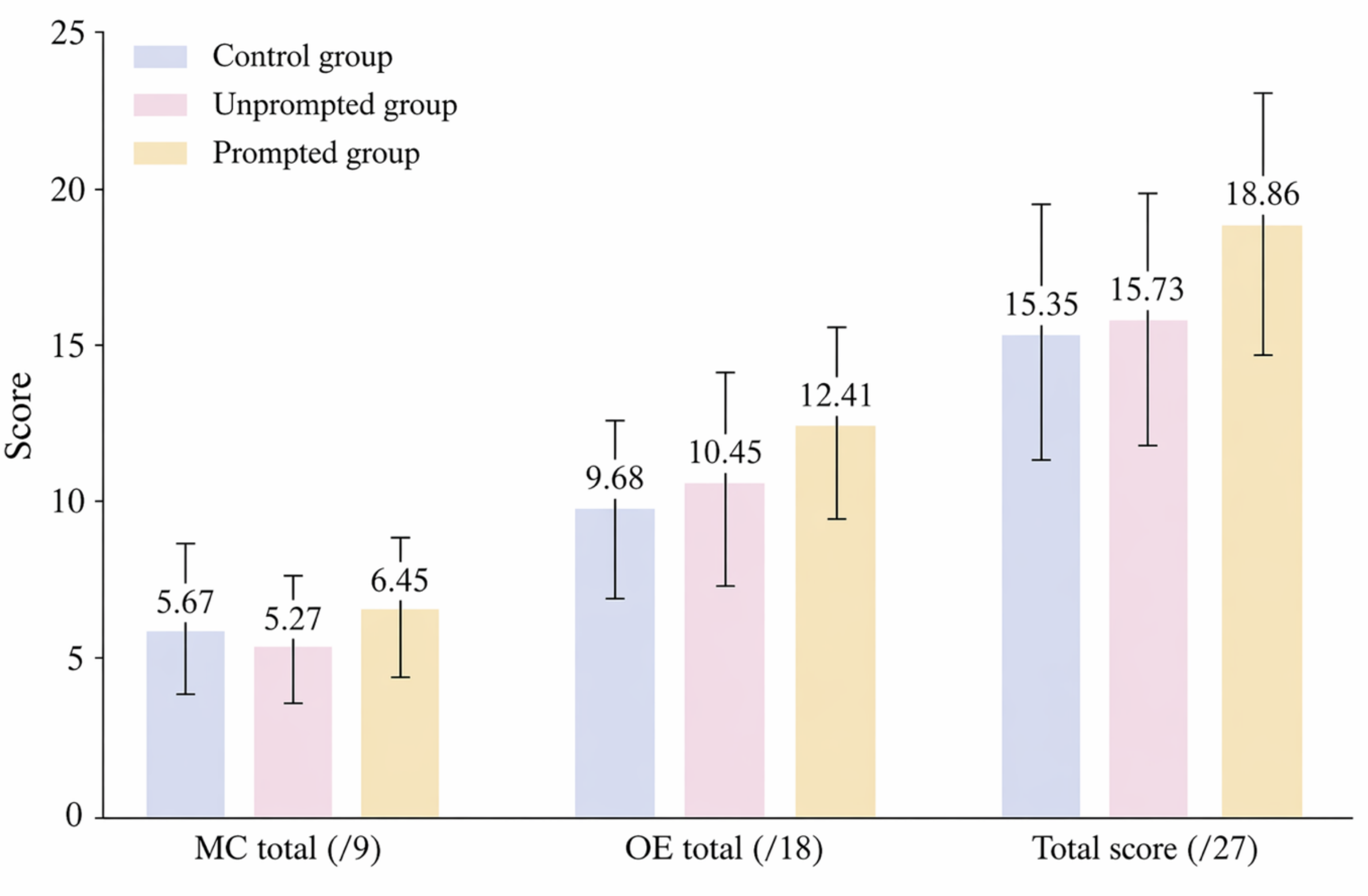

**Fig. 6.** Learning performance across the three conditions (mean ± SD)

**Table 3.** Post-review learning performance (item-level means)

| Item | Control group | Unprompted group | Prompted group | F-value | p-value | Significance |
|---|---|---|---|---|---|---|
| **Multiple-choice questions (max = 1)** | | | | | | |
| Question 1 | 0.55 | 0.45 | 0.69 | - | - | - |
| Question 2 | 0.88 | 0.97 | 0.90 | - | - | - |
| Question 3 | 0.85 | 0.79 | 0.90 | - | - | - |
| Question 4 | 0.48 | 0.27 | 0.52 | - | - | - |
| Question 5 | 0.55 | 0.64 | 0.72 | - | - | - |
| Question 6 | 0.52 | 0.45 | 0.69 | - | - | - |
| Question 7 | 0.73 | 0.48 | 0.72 | - | - | - |
| Question 8 | 0.55 | 0.58 | 0.66 | - | - | - |
| Question 9 | 0.58 | 0.64 | 0.66 | - | - | - |
| **MC accuracy (mean ± SD)** | 0.63 ± 0.26 | 0.59 ± 0.22 | 0.72 ± 0.24 | - | - | - |
| **Mean total (mean ± SD)** | 5.67 ± 2.31 | 5.27 ± 1.96 | 6.45 ± 2.15 | 2.38 | 0.099 | ns |
| **Open-ended questions (max = 6)** | | | | | | |
| Question 10 | 3.67 | 3.95 | 4.40 | - | - | - |
| Question 11 | 3.03 | 3.03 | 4.26 | - | - | - |
| Question 12 | 2.98 | 3.47 | 3.76 | - | - | - |
| **OE accuracy (mean ± SD)** | 3.23 ± 0.80 | 3.48 ± 1.11 | 4.14 ± 0.93 | - | - | - |
| **OE total (mean ± SD)** | 9.68 ± 2.39 | 10.45 ± 3.34 | 12.41 ± 2.79 | 7.32 | 0.0011 | ** |
| **Total score (mean ± SD)** | 15.35 ± 4.11 | 15.73 ± 4.06 | 18.86 ± 4.01 | 6.82 | 0.0017 | ** |

*Note.* SD = standard deviation.

For multiple-choice questions, the Prompted group achieves the highest mean score (6.45), followed by the NA (5.67) and Unprompted (5.27) groups. However, the ANOVA results indicate that this difference is not statistically significant (F = 2.38, p = 0.099). This finding suggests that the three learning conditions yield comparable outcomes in lower-level cognitive tasks, such as factual recall and recognition, implying that basic knowledge acquisition is not substantially influenced by AI use or prompting strategies.

In contrast, a significantly different pattern is observed for open-ended questions. The Prompted group achieves a substantially higher mean score (12.41) than the NA (9.68) and Unprompted (10.45) groups, with the difference reaching statistical significance (F = 7.32, p = 0.0011). This pattern is clearly reflected in Figure 6, where the Prompted condition shows a noticeable advantage over the other two groups. Given that open-ended tasks require explanation, reasoning, and knowledge integration, this result indicates that structured prompting is associated with improved performance in higher-order cognitive processes. Notably, the absence of a comparable improvement in the Unprompted condition suggests that the observed benefit is not attributable to AI usage alone, but to the manner in which learners are guided to interact with the system.

A similar pattern is observed in the overall scores. The Prompted group achieves the highest mean performance (18.86), significantly exceeding both the NA (15.35) and Unprompted (15.73) groups ($F = 6.82$, $p = 0.0017$). This result indicates that the effect of the prompting strategy extends beyond individual task types and contributes to overall learning performance.

The results indicate that theory-guided prompting has limited impact on basic knowledge recall but plays a significant role in supporting higher-order learning outcomes. This pattern is consistent with the UEQ results, where improvements are primarily observed in pragmatic dimensions such as Perspicuity and Dependability. One possible explanation is that increased interaction clarity and perceived reliability reduce cognitive load and facilitate more structured reasoning processes, thereby supporting performance in open-ended tasks.

## DISCUSSION

### Effects of Prompting on Learning Outcomes

Differences across the three learning conditions, namely conventional slide-based learning (NA), unprompted AI-supported learning, and prompted AI-supported learning, are primarily observed in tasks requiring explanation and reasoning. Performance on multiple-choice questions remains comparable, indicating similar effectiveness in recognition-based tasks regardless of the use of structured prompting. These tasks mainly involve identifying and distinguishing key concepts and do not require extensive cognitive processing, which limits the contribution of additional instructional support.

In contrast, higher performance is observed under the prompted condition for open-ended responses. Tasks involving explanation and knowledge integration appear to benefit from a more structured interaction process, through which learners clarify key ideas, organize relationships, and refine their understanding. This process supports the development of more coherent responses and reduces ambiguity in reasoning. The lack of a comparable improvement in the unprompted condition indicates that access to AI alone is insufficient to produce this effect; rather, the structure of the interaction plays a central role in shaping how learners engage with the material.

**Relationship Between User Experience and Performance**

The UEQ results and the performance outcomes show a consistent pattern across conditions, although the relationship is not uniform across all dimensions. Differences in performance are mainly observed in tasks that require explanation and reasoning, and these differences coincide with variations in how learners perceive the clarity and reliability of the interaction.

In the prompted condition, higher scores in Perspicuity and Dependability correspond to stronger performance in open-ended tasks. The interaction process in this condition is more explicit, which appears to make it easier for learners to follow the logic of the response and to structure their own explanations. This becomes more relevant when tasks require the integration of multiple concepts rather than the identification of correct answers. This alignment is less evident for Efficiency. Perceived efficiency remains similar across conditions, while both time expenditure and performance vary. Faster interaction therefore does not necessarily correspond to better outcomes, particularly in tasks where additional processing is required. A similar disconnect can be observed for Stimulation and Novelty. These dimensions vary only slightly and do not follow the same pattern as performance. Although they reflect aspects of engagement, their role in supporting task completion appears limited in this context.

The pattern that emerges is that performance differences tend to appear alongside changes in interaction clarity and structure, rather than alongside differences in engagement or perceived efficiency. This pattern is consistent with the nature of the tasks, which emphasize reasoning and explanation over speed or exploration interaction.

**Implications for AI-supported Learning Design**

The findings highlight the importance of structuring student–AI interaction to support cognitive processing. The absence of measurable improvement in the unprompted AI-supported condition indicates that simple access to AI does not inherently lead to better learning outcomes. In contrast, the prompted condition demonstrates that guided interaction can shape how learners engage with instructional content. When learners are directed through a sequence of clarification, organization, and verification, their engagement becomes more systematic and aligned with the demands of the learning task. This suggests that the effectiveness of AI-supported learning environments depends on how interaction is organized and guided.

From a design perspective, prompting can be understood as a form of instructional scaffolding embedded within the interaction process. The prompting framework directs attention, supports the organization of information, and encourages reflective evaluation of understanding. Through these mechanisms, AI functions as a structured learning support that influences the way learners process information. Designing AI systems with explicit interaction structures may therefore be essential for supporting higher-order learning outcomes, particularly in contexts that require explanation, reasoning, and conceptual understanding.

**Limitations**

Several limitations should be acknowledged when interpreting the findings. First, the study is conducted within a single instructional context with a specific set of learning materials, which may limit the generalizability of the results to other domains or course formats. Second, the sample size is moderate, and the group sizes are not fully balanced due to data exclusion, which may affect the stability of statistical comparisons. Third, the evaluation is based on a short-term learning task, and the results do not capture longer-term retention or transfer of knowledge. In addition, time expenditure is analyzed descriptively without formal statistical testing, which limits the strength of conclusions regarding efficiency-related differences.

The design of the prompting framework also introduces certain constraints. The structured sequence may not accommodate all individual learning preferences, and some learners may benefit from more flexible interaction patterns. Furthermore, the study focuses on a single prompting strategy without comparison to alternative designs, leaving open the question of how different forms of guidance may influence learning. Future work can extend this research by examining diverse instructional contexts, incorporating longitudinal measures, and exploring adaptive prompting mechanisms that respond to learner characteristics and task demands.

## CONCLUSION

This study examines how instructional guidance, implemented through structured prompting shapes student–AI interaction during self-directed learning in construction education. A five-step prompting framework grounded in Generative Learning Theory is proposed and evaluated through a controlled comparison with conventional slide-based learning and unprompted AI-supported learning. The results show that the prompted condition consistently improves performance in tasks that require explanation and reasoning, while no measurable difference is observed in recognition-based tasks. This pattern indicates that the primary effect of instructional guidance lies in supporting higher-order cognitive processes.

The findings further show that the effect of instructional guidance is associated with changes in how learners engage with the task. The prompted condition requires more time to complete the learning activity, reflecting a more deliberate and structured interaction process. This additional time investment is not accompanied by improvements in perceived efficiency, yet it is associated with stronger performance in open-ended responses. Overall, these results point to a shift in learning behavior, where instructional guidance, operationalized through structured prompting, supports deeper cognitive processing and more coherent knowledge construction. The absence of similar improvement in the unprompted condition suggests that access to AI alone is insufficient, and that the organization of interaction plays a central role in shaping learning outcomes.

From a design perspective, the study positions instructional guidance as an instructional mechanism embedded within AI-supported learning environments. The proposed framework provides a structured approach to guiding learner interaction, supporting processes of clarification, organization, and verification that are essential for conceptual understanding. Several limitations should be considered when interpreting the findings. The study is conducted within a single instructional context with a moderate sample size, and the evaluation focuses on short-term performance without addressing long-term retention. In addition, time differences are described at a descriptive level without formal statistical testing. Future work can extend this research by examining diverse learning contexts, incorporating longitudinal measures, and exploring adaptive forms of instructional guidance that respond to learner characteristics.

**ACKNOWLEDGEMENTS**

This work is supported by the National Science Foundation (NSF), award # 2417804. Any opinions, findings, and conclusions or recommendations expressed in this paper are those of the author(s) and do not necessarily reflect the views of the NSF.

## DATA AVAILABILITY STATEMENT

Some or all data, models, or code that support the findings of this study are available from the corresponding author upon reasonable request.